\documentclass[showkeys,showpacs,superscriptaddress]{revtex4}
\usepackage[caption=false]{subfig}
\usepackage{amsmath}
\usepackage{amssymb}
\usepackage{graphicx}
\usepackage{bm}

\begin{document}

\title{
Monopole solutions in SU(2) Yang-Mills-plus-massive-nonlinear-spinor-field theory
}

\author{
Vladimir Dzhunushaliev
}
\email{v.dzhunushaliev@gmail.com}
\affiliation{
Department of Theoretical and Nuclear Physics,  Al-Farabi Kazakh National University, Almaty 050040, Kazakhstan
}
\affiliation{
Institute of Experimental and Theoretical Physics,  Al-Farabi Kazakh National University, Almaty 050040, Kazakhstan
}
\affiliation{
Academician J.~Jeenbaev Institute of Physics of the NAS of the Kyrgyz Republic, 265 a, Chui Street, Bishkek 720071, Kyrgyzstan
}

\author{Vladimir Folomeev}
\email{vfolomeev@mail.ru}
\affiliation{
Institute of Experimental and Theoretical Physics,  Al-Farabi Kazakh National University, Almaty 050040, Kazakhstan
}
\affiliation{
Academician J.~Jeenbaev Institute of Physics of the NAS of the Kyrgyz Republic, 265 a, Chui Street, Bishkek 720071, Kyrgyzstan
}

\author{Albina Serikbolova}
\email{albeni_23_95@mail.ru}
\affiliation{
Department of Theoretical and Nuclear Physics,  Al-Farabi Kazakh National University, Almaty 050040, Kazakhstan
}


\begin{abstract}
Monopole solutions in SU(2) Yang-Mills theory which includes spinor fields described by the nonlinear Dirac equation are obtained.
It is demonstrated that the energy spectrum of such a system possesses a global minimum whose appearance
is brought about solely by the nonlinear spinor fields. It is shown that the monopole solution obtained differs in principle from the 't~Hooft-Polyakov monopole in that it is topologically trivial.
\end{abstract}

\pacs{12.38.Mh, 11.15.Tk, 12.38.Lg, 11.15.-q
}

\keywords{
non-Abelian SU(2) theory; nonlinear Dirac equation; monopole;  energy spectrum
}
\date{\today}

\maketitle

\section{Introduction}

In recent decades magnetic monopoles obtained within the framework of non-Abelian Yang-Mills theories find their applications to a wide variety of topics in theoretical physics,
including various problems in the standard model and its extensions, astrophysics, and cosmology~\cite{Shnir:2005}.
The simplest example of a regular localized monopole solution in SU(2) Yang-Mills theory is the well-known 't~Hooft-Polyakov monopole~\cite{tHooft:1974kcl,Polyakov:1974ek}.
The obligatory condition for its existence is the inclusion into the system of a triplet of Higgs scalar fields ensuring the presence of a topological charge.
A distinctive feature of such scalar fields is their nontrivial behaviour at spatial infinity. In this connection, one might suppose that, if it would be possible to find regular
monopole solutions without involving scalar fields, then they might already be topologically trivial. Consistent with this,
the main purpose of the present Letter is to demonstrate the possibility of the existence of monopole solutions without scalar fields, which are replaced by a nonlinear spinor field.

The study of nonlinear spinor fields was initiated by W.~Heisenberg in the 1950's. His main idea was the assumption that the nonlinear Dirac equation
can describe the internal structure of an electron. In other words, this equation  is a fundamental equation which enables one to get all main characteristics
of an electron: its spin, charge, and mass. However, with the advent of quantum electrodynamics, further investigations in this direction were discontinued;
one of the reasons for that was that the theory based on the nonlinear Dirac equation is nonrenormalizable. Next time the nonlinear Dirac equation has
appeared as applied to an  approximate description of hadrons within the Nambu-Jona-Lasinio model \cite{Nambu:1961tp}  (for a review, see Ref.~\cite{Klevansky:1992qe}).
In that model, a nonlinear spinor field is not fundamental but is used as some approximation within QCD. Notice also that, unlike the Nambu-Jona-Lasinio model,
 in the present work we study the nonlinear Dirac equation with a mass term.

In this Letter, we study topologically trivial monopole solutions in SU(2) Yang-Mills theory containing also a spinor field described by the nonlinear Dirac equation.
To do so, we seek monopole solutions with the source of  magnetic field in the form of color charge created by the spinor field. We study here the energy spectrum of a spherically
symmetric system consisting of a SU(2) Yang-Mills field and a nonlinear spinor field. We show that in such a system there exist regular solutions possessing a finite energy
whose energy spectrum has a minimum. The reason for the appearance of such a minimum is the presence in the system of a nonlinear Dirac field.
In this connection, let us notice that the energy spectrum of regular solutions  for a system supported only by a nonlinear spinor field was obtained
in the pioneering works~\cite{Finkelstein:1951zz,Finkelstein:1956} where it was shown that this spectrum has a minimum corresponding to ``the lightest stable particle.''

The Letter is organized as follows. In Sec.~\ref{YM_Dirac_scalar}, we write down the Lagrangian and general field equations for SU(2) Yang-Mills theory
containing a nonlinear spinor field. In Sec.~\ref{monopole_Dirac}, we present the {\it Ans\"{a}tze} for vector and spinor fields and also the corresponding equations.
Numerical solutions to these equations are sought in Sec.~\ref{monopole_Sols}, while in Sec.~\ref{MassGap}
we study their energy spectrum and show the presence of global minima.

\section{Theory of Yang-Mills fields coupled to a nonlinear Dirac field}
\label{YM_Dirac_scalar}

The Lagrangian describing a system consisting of a non-Abelian SU(2) field $A^a_\mu$ interacting with nonlinear spinor field $\psi$ can be taken in the form
\begin{equation}
\begin{split}
	\mathcal L = & - \frac{1}{4} F^a_{\mu \nu} F^{a \mu \nu}
	+ i \hbar c \bar \psi \gamma^\mu D_\mu \psi  -
	m_f c^2 \bar \psi \psi+
	\frac{\Lambda}{2} g \hbar c \left( \bar \psi \psi \right)^2.
\label{1_10}
\end{split}
\end{equation}
Here $m_f$ is the mass of the spinor field;
$
D_\mu = \partial_\mu - i \frac{g}{2} \sigma^a
A^a_\mu
$ is the gauge-covariant derivative, where $g$ is the coupling constant and $\sigma^a$ are the SU(2) generators (the Pauli matrices);
$
F^a_{\mu \nu} = \partial_\mu A^a_\nu - \partial_\nu A^a_\mu +
g \epsilon_{a b c} A^b_\mu A^c_\nu
$ is the field strength tensor for the SU(2) field, where $\epsilon_{a b c}$ (the completely antisymmetric Levi-Civita symbol)
are the SU(2) structure constants;  $\Lambda$ is a constant; $\gamma^\mu$ are the Dirac matrices in the standard representation;
$a,b,c=1,2,3$ are color indices and $\mu, \nu = 0, 1, 2, 3$ are spacetime indices.

Using Eq.~\eqref{1_10}, the corresponding field equations can be written in the form
\begin{eqnarray}
	D_\nu F^{a \mu \nu} &=& \frac{g \hbar c}{2}
	\bar \psi \gamma^\mu \sigma^a \psi ,
	\label{1_20}\\
	i \hbar \gamma^\mu D_\mu \psi  - m_f c \psi + \Lambda g \hbar \psi
	\left(
		\bar \psi \psi
	\right)&=& 0.
\label{1_30}
\end{eqnarray}
Let us enumerate some distinctive features of the system under consideration:
(i)~The set of equations~\eqref{1_20} and \eqref{1_30} has
monopole-like solutions only for some special choices of the system parameters $f_2$ and $u_1$ [for their definition see Eq.~\eqref{T_series}];
(ii)~In the absence of the vector field $A^a_\mu$, there exist particlelike solutions of the nonlinear Dirac equation  \eqref{1_30}
which describe a system possessing a minimum in the energy spectrum~\cite{Finkelstein:1951zz,Finkelstein:1956};
(iii)~In the absence of the spinor field, the Yang-Mills equation~\eqref{1_20} has no static globally regular solutions~\cite{Deser:1976wq};
(iv)~To the best of our knowledge,
in the case of linear spinor field (i.e., when $\Lambda=0$) the set of equations~\eqref{1_20} and \eqref{1_30} has no static regular solutions as well.

To obtain particlelike solutions, Eqs.~\eqref{1_20} and \eqref{1_30} will be solved numerically as an eigenvalue problem for the parameters
$f_2$ and $u_1$, since apparently it is impossible to find their analytical solution.

\section{ Ans\"{a}tze and equations }
\label{monopole_Dirac}

We seek monopole-like solutions to Eqs.~\eqref{1_20} and \eqref{1_30} describing objects consisting of a radial magnetic field and a nonlinear spinor field. For this purpose, we employ the standard SU(2) monopole {\it Ansatz}
\begin{eqnarray}
	A^a_i &=&  \frac{1}{g} \left[ 1 - f(r) \right]
	\begin{pmatrix}
		0 & \phantom{-}\sin \varphi &  \sin \theta \cos \theta \cos \varphi \\
		0 & -\cos \varphi &   \sin \theta \cos \theta \sin \varphi \\
		0 & 0 & - \sin^2 \theta
	\end{pmatrix} , \quad
\nonumber \\
	&&
	i = r, \theta, \varphi  \text{ (in polar coordinates)},
\label{2_10}\\
	A^a_t &=& 0 ,
\label{2-13}
\end{eqnarray}
and the {\it Ansatz} for the spinor field from Refs.~\cite{Li:1982gf,Li:1985gf}
\begin{equation}
	\psi^T = \frac{e^{-i \frac{E t}{\hbar}}}{g r \sqrt{2}}
	\begin{Bmatrix}
		\begin{pmatrix}
			0 \\ - u \\
			\end{pmatrix},
			\begin{pmatrix}
			u \\ 0 \\
			\end{pmatrix},
			\begin{pmatrix}
			i v \sin \theta e^{- i \varphi} \\ - i v \cos \theta \\
			\end{pmatrix},
			\begin{pmatrix}
			- i v \cos \theta \\ - i v \sin \theta e^{i \varphi} \\
		\end{pmatrix}
	\end{Bmatrix},
\label{2_20}
\end{equation}
where $E/\hbar$ is the spinor frequency and the functions $u$ and $v$ depend on the radial coordinate $r$ only. In Eq.~\eqref{2_20}, each row describes a spin-$1/2$ fermion, and these two fermions have the same mass $m_f$ and opposite spins and are located at one point. Aside from this, for each of such fermions, the energy-momentum tensors will not be spherically symmetric (due to the existence of nondiagonal components), but their sum will give a tensor compatible with spherical symmetry of the system under consideration.

Equations for the unknown functions $f, u$, and $v$ can be obtained  by substituting the expressions~\eqref{2_10}-\eqref{2_20} into the field equations~\eqref{1_20} and \eqref{1_30},
\begin{eqnarray}
	- f^{\prime \prime} + \frac{f \left( f^2 - 1 \right) }{x^2} +
	\tilde g^2\,\frac{\tilde u \tilde v}{x} &=& 0 ,
\label{2_30}\\
	\tilde v' + \frac{f \tilde v}{x} &=& \tilde u \left(
		- \tilde m_f + \tilde E +
		\tilde \Lambda \frac{\tilde u^2 - \tilde v^2}{x^2}
	\right) ,
\label{2_40}\\
	\tilde u' - \frac{f \tilde u}{x} &=& \tilde v \left(
		- \tilde m_f - \tilde E +
		\tilde \Lambda   \frac{\tilde u^2 - \tilde v^2}{x^2}
	\right).
\label{2_50}
\end{eqnarray}
Here, for convenience of making numerical calculations, we have introduced the following dimensionless variables:
$x = r/r_0$, where $r_0$ is a constant corresponding to the characteristic size of the system under consideration;
$\tilde u=\sqrt{r_0}u/g,
\tilde v = \sqrt{r_0}v/g,
\tilde m_f = r_0 m_f c/\hbar,
\tilde E = r_0 E/(\hbar c),
\tilde \Lambda = \left(g /r_0^2\right)\Lambda
$, $\tilde g^2 = g^2 \hbar c$.
The prime denotes differentiation with respect to  $x$.
The parameter $r_0$ must depend only on constants of a theory;
therefore one can take, say, $r_0 = \hbar/(m_f c)$.

The total energy density of the monopole-plus-spinor-fields system under consideration is
\begin{equation}
	\tilde \epsilon =
	\tilde{\epsilon}_m + \tilde \epsilon_s =\frac{1}{\tilde g^2}
	\left[
		\frac{{f'}^2}{ x^2} +
		\frac{\left( f^2 - 1 \right)^2}{2 x^4}
	\right] +
	\left[
		\tilde E \frac{\tilde u^2 + \tilde v^2}{x^2} +
		\frac{\tilde\Lambda}{2}
		\frac{\left(\tilde u^2 - \tilde v^2 \right)^2}{x^4}
	\right],
\label{2_60}
\end{equation}
where the expressions in the square brackets correspond to
the dimensionless energy densities of the monopole, $\tilde{\epsilon}_m\equiv \left(r_0^4/\hbar c\right)\epsilon_m$,
and of the spinor field, $\tilde{\epsilon}_s\equiv \left(r_0^4/\hbar c\right)\epsilon_s$.

\section{Monopole-plus-spinor-fields solutions}
\label{monopole_Sols}

This section is devoted to the numerical study of monopole-plus-spinor-fields solutions of Eqs.~\eqref{2_30}-\eqref{2_50}. Because of the presence
of the terms containing $x$ in the denominators of these equations, to perform numerical computations, we assign boundary conditions near the origin
 $x = 0$  where solutions are sought in the form of the Taylor series
\begin{equation}
f = 1 + \frac{f_2}{2} x^2 + \ldots ,\quad
	\tilde u = \tilde u_1 x + \frac{\tilde u_3}{3!} x^3 + \ldots ,\quad
	\tilde v = \frac{\tilde v_2}{2} x^2 + \frac{\tilde v_4}{4!} x^4 + \ldots ,
\label{T_series}
\end{equation}
where $\tilde v_2 = 2 \tilde u_1 \left(
		\tilde E-\tilde m_f + \tilde\Lambda \tilde u_1^2
	\right)/3$
and the expansion coefficients $f_2$ and $\tilde u_1$  are free parameters whose values cannot be found from Eqs.~\eqref{2_30}-\eqref{2_50}.

\begin{figure}[t]
\centering
		\includegraphics[width=1\linewidth]{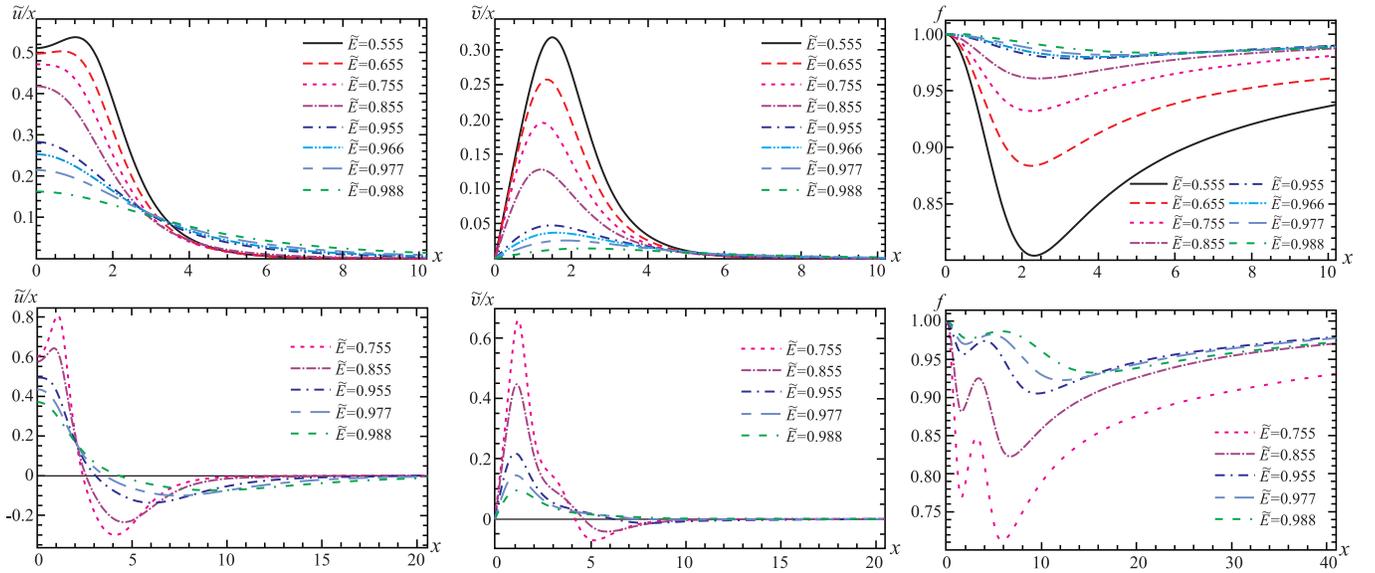}
	\caption{The functions $\tilde u(x)/x$, $\tilde v(x)/x$, and $f(x)$ for different values of the parameter
$\tilde E$ with $\tilde \Lambda=8$, $\tilde m_f=1$, and $\tilde g=1$. The figures in the top row show the solutions for the ground state of the system, and those in the bottom row~-- for the first excited state.
	}
	\label{fields_x}
\end{figure}

Eqs.~\eqref{2_30}-\eqref{2_50} are solved numerically as a nonlinear problem for the eigenvalues $f_2$ and $\tilde u_1$ and the eigenfunctions $\tilde u, \tilde v,$ and $f$,
whose typical behavior is shown in Fig.~\ref{fields_x} both for the ground state of the system under consideration and for the first excited state when the functions $\tilde u$ and $\tilde v$ possess one node.
The corresponding computed values of the system parameters are given in Table~\ref{MG_monopole}.

\begin{figure}[h!]
\centering
		\includegraphics[width=1\linewidth]{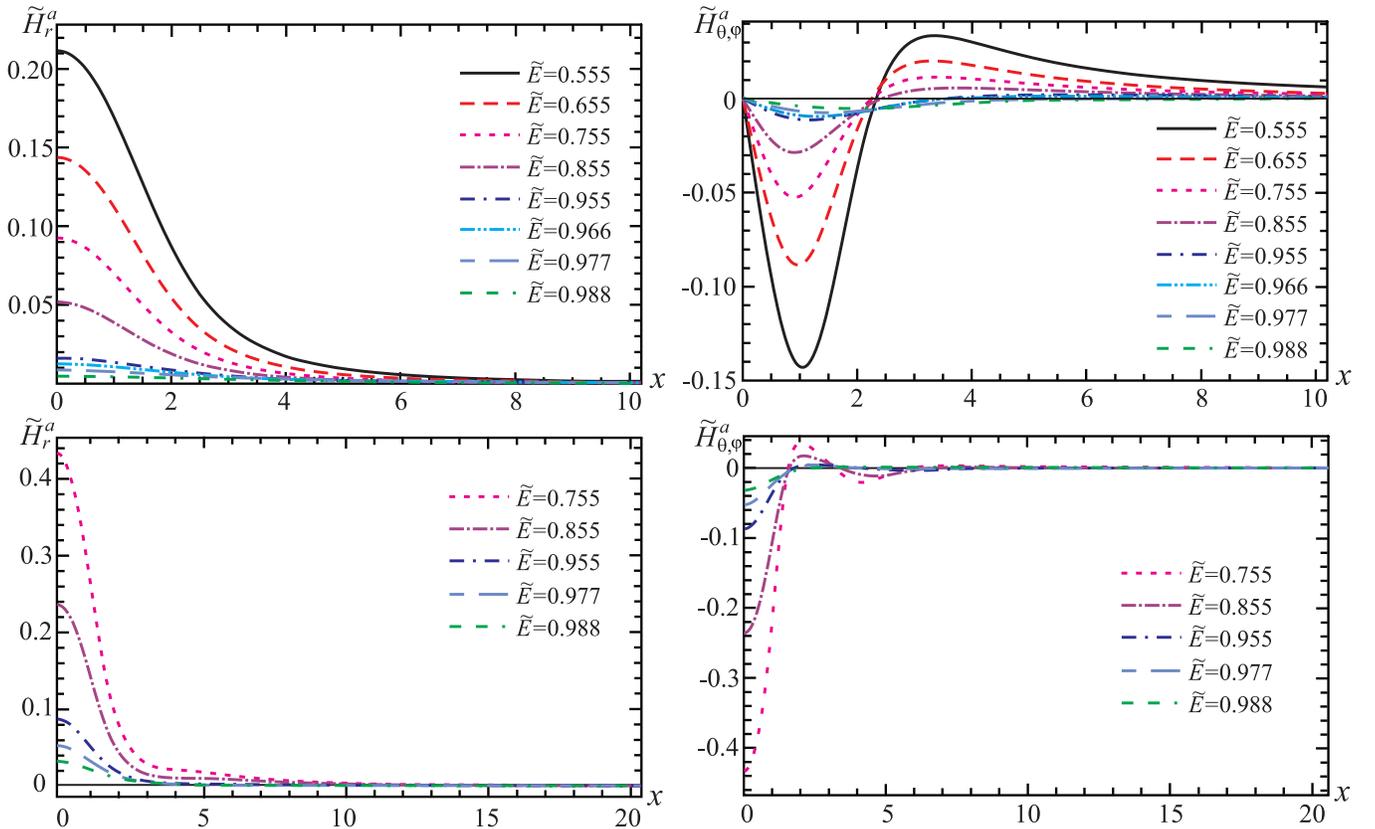}
	\caption{The distributions of the color magnetic fields for different values of the parameter $\tilde E$: the radial component $\tilde H^a_r\equiv g r_0^2H^a_r$ is given by Eq.~\eqref{3_70}
and the tangential components $\tilde H^a_{\theta, \varphi}\equiv g r_0 H^a_{\theta, \varphi}$~-- by Eq.~\eqref{3_81}.
The figures in the top row correspond to the ground state of the system, and those in the bottom row~-- to the first excited state.
	}
	\label{magn_x}
\end{figure}

\begin{table}[h!]
\scalebox{0.68}{
\begin{tabular}{|c|c|c|c|c|c|c|c|c|c|}
	\hline
			\multicolumn{9}{|c|}{The ground state} \\
	\hline
	\rule[-1ex]{0pt}{2.5ex}
	$\tilde E$&0.555&0.655&0.755&0.855&0.955&0.966&0.977&0.988\\
	\hline
	\rule[-1ex]{0pt}{2.5ex}
	$f_2$&-0.21167&-0.1438&-0.092587&-0.0519&-0.016338&-0.012473&
		-0.0085451&-0.0046615	\\
	\hline
	\rule[-1ex]{0pt}{2.5ex}
	$\tilde u_1$&0.510757372&0.497238675&0.47145607&0.4184815&0.2834&0.25342&
	0.2151&0.163						\\
	\hline
	\rule[-1ex]{0pt}{2.5ex}
	$\tilde W_t$&15.6339&11.7187&8.6202&6.4621&5.8124&6.049896&6.5242&7.3827\\
	\hline
			\multicolumn{9}{|c|}{The first excited state, one-node solutions} \\
	\hline
	\rule[-1ex]{0pt}{2.5ex}
	$\tilde E$&&&0.755&0.855&0.955&0.977&0.988&\\
	\hline
	\rule[-1ex]{0pt}{2.5ex}
	$f_2$&&&-0.43377&-0.23663&-0.087295&-0.0526425&-0.03217&\\
\hline
	\rule[-1ex]{0pt}{2.5ex}
	$\tilde u_1$&&&0.6005&0.57331012&0.494131&0.43494&0.37143&\\
	\hline
	\rule[-1ex]{0pt}{2.5ex}
	$\tilde W_t$&&&76.182&62.582&53.748&57.3803&65.957&\\
	\hline
\end{tabular}
}
\caption{
	Eigenvalues $\tilde u_1$ and $f_2$ and the total energy $\tilde W_t$ from Eq.~\eqref{4_10} for different values of the parameter $\tilde E$.
}
\label{MG_monopole}
\end{table}

Asymptotically (as $x \to \infty$), the behavior of the solutions is
$$f(x) \approx 1 - \frac{f_\infty}{x} , \quad
	\tilde u(x) \approx \tilde u_\infty
	e^{- x \sqrt{\tilde m_f^2 - \tilde E^2}} ,\quad
	\tilde v(x) \approx
	\tilde v_\infty e^{- x \sqrt{\tilde m_f^2 - \tilde E^2}} ,$$
where $f_\infty, \tilde u_\infty$,  and $\tilde v_\infty$ are integration constants.

It is of interest to follow the behavior of the magnetic Yang-Mills  field. Its physical components can be defined as
$
H^a_i=-(1/2)\sqrt{\gamma}\,\epsilon_{i j k} F^{a j k},
$
where $i, j, k$ are space indices. In our case this gives for the radial magnetic field
\begin{equation}
		H^a_r \sim \frac{1 - f^2}{g r^2},
\label{3_70}
\end{equation}
where $a=1,2,3$ and we have dropped the dependence on the angular variables.
The corresponding graphs for this component are shown in Fig.~\ref{magn_x}. In turn, its asymptotic behavior as $x\to \infty$ is
\begin{equation}
	H^a_r \sim \frac{2 f_\infty}{g r^3}.
\label{3_80}
\end{equation}
It is seen from this expression that, by its asymptotic behavior, the system monopole-plus-nonlinear-spinor-fields differs in principle
from the 't~Hooft-Polyakov monopole, whose magnetic field decreases as $r^{-2}$.

 Nonzero tangential components of the magnetic field are
\begin{equation}
H^a_{\theta}\sim\frac{1}{g}f^{\prime}, \quad H^b_{\varphi}\sim\frac{1}{g}f^{\prime},
\label{3_81}
\end{equation}
where $a=1,2,3$ and $b=1,2$.
Their behavior is shown in Fig.~\ref{magn_x}.

Thus in this section we have obtained the spherically symmetric solutions describing the self-consistent system consisting of the non-Abelian magnetic field and non-linear spinor field. Let us emphasize the important feature of the monopole described by Eqs.~\eqref{2_30}-\eqref{2_50}: this monopole is topologically trivial, since for its existence the presence of a scalar field triplet, whose behavior at spatial infinity is topologically nontrivial, is not needed.

\section{Energy spectrum }
\label{MassGap}

\begin{figure}[t]
\centering
		\includegraphics[width=1\linewidth]{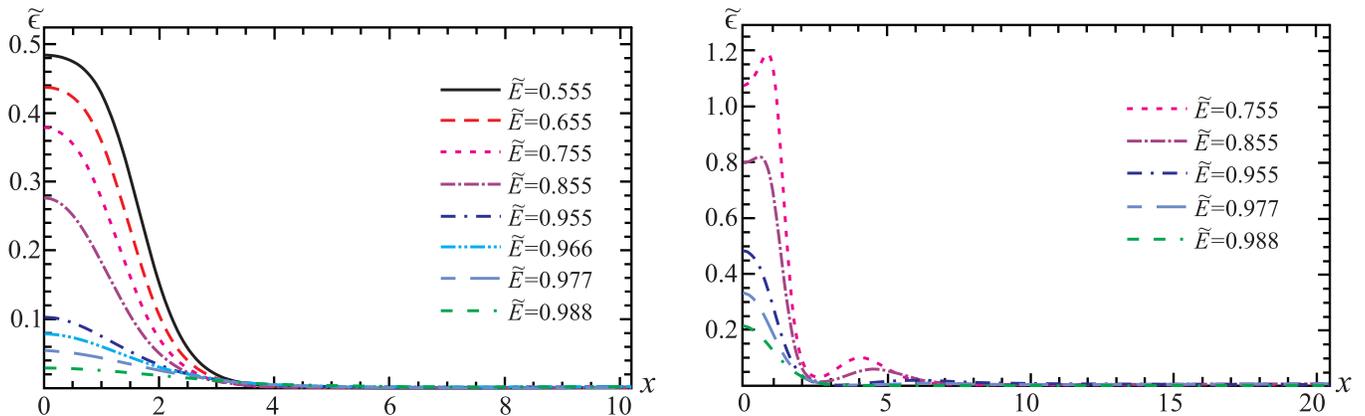}
	\caption{The energy density $\tilde \epsilon$ from Eq.~\eqref{2_60} for different values of the parameter $\tilde E$.
The left panel corresponds to the ground state of the system, and the right one~-- to the first excited state.
	}
	\label{energy_density}
\end{figure}

\begin{figure}[t]
\centering
		\includegraphics[width=.5\linewidth]{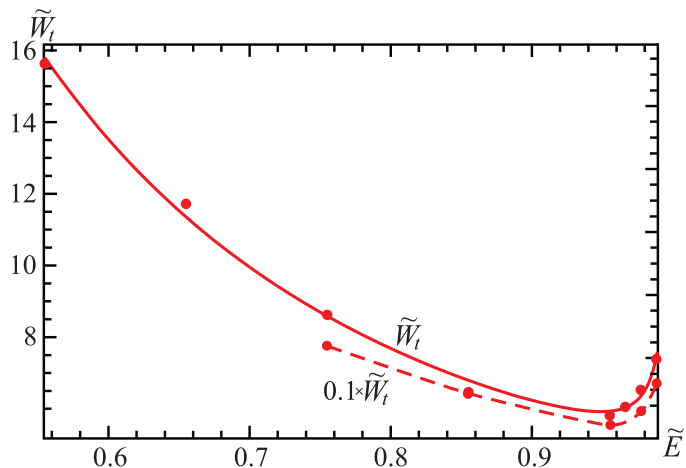}
	\caption{The spectrum of the total energy for the ground (solid line) and excited (dashed line) states
	from Eq.~\eqref{4_20} as functions of the parameter $\tilde E$ (the bold dots show the values of $\tilde W_t$ taken from Table~\ref{MG_monopole}).
	}
	\label{energy_spectrum}
\end{figure}

In this section we obtain the energy spectrum of the configuration under consideration as a function of the parameter $\tilde E$
and demonstrate the presence of a minimum in the spectrum. For this purpose, we employ an expression for a dimensionless total energy
of the system in question,
 \begin{equation}
	\tilde W_t \equiv \frac{W_t}{ \hbar c/r_0  } = 4 \pi
	\int\limits_0^\infty x^2 \tilde \epsilon d x =
	\left( \tilde{W}_t \right)_m + \left( \tilde{W}_t \right)_{s},
\label{4_10}
\end{equation}
where the energy density $\tilde \epsilon$ is taken from Eq.~\eqref{2_60}.
One can see from this formula that the total energy is split into a sum of energies of the monopole, $\left( \tilde{W}_t \right)_m$, and of the spinor fields, $\left( \tilde{W}_t \right)_{s}$, despite the presence of the direct interaction between the vector and spinor fields. The corresponding distributions of $\tilde \epsilon$ along the radius are shown in Fig.~\ref{energy_density}.
In turn, using Eq.~\eqref{4_10}, we have calculated the magnitudes of the total energy given in Table~\ref{MG_monopole}. Using them, we have plotted in
Fig.~\ref{energy_spectrum} the corresponding energy spectrum of the system.

 The calculated data for $\tilde W_t$ given in Table~\ref{MG_monopole} can be interpolated by the fitting formula
\begin{equation}
	[\tilde W_t]_{\text{fit}} = a\tilde E^\alpha +
	b\left( 1 - \tilde E \right)^\beta .
\label{4_20}
\end{equation}
In particular, for the ground state, $a = 4.80$, $\alpha = -2.02$, $b = 0.03$, and $\beta = -1.01$. Using this formula, in Fig.~\ref{energy_spectrum},
we have plotted the curve that illustrates the existence of a minimum in the energy spectrum of the monopole solutions obtained.

Thus in this section we have obtained the energy spectrum of the self-consistent system modeled within SU(2) Yang-Mills theory containing the doublet of nonlinear spinor fields.
The important result of the calculations is that the energy spectrum possesses a global minimum (see Fig.~\ref{energy_spectrum}).

\section{Summary}
\label{concl}

Thus, we have found topologically trivial monopole solutions within SU(2) Yang-Mills theory containing a doublet of nonlinear spinor fields.
These solutions describe a magnetic monopole created by a spherical lump of nonlinear spinor fields. This monopole differs in principle from the 't~Hooft-Polyakov
monopole in that it is topologically trivial. The solutions obtained here do not exist without the spinor fields which are the source of the color magnetic field.
This enables us to arrive at an important conclusion that the reason for the appearance of a minimum in the energy spectrum is the presence of the nonlinear Dirac fields.
This conclusion is also confirmed by the results of Refs.~\cite{Finkelstein:1951zz,Finkelstein:1956} where a similar minimum was found for the energy spectrum of regular
solutions to the nonlinear Dirac equation (the authors called it ``the lightest stable particle'').

Summarizing the results obtained,
\begin{itemize}
\item We have found regular finite-energy monopole solutions with a colour magnetic field sourced by a nonlinear Dirac field.
\item It is demonstrated that the energy spectrum has a global minimum, both for the ground state and for the first excited state.
\item  It is shown that the main reason for the appearance of the minimum in the energy spectrum
is the presence of the nonlinear spinor fields.
\end{itemize}




\begin{thebibliography}{00}



\bibitem{Shnir:2005}
Ya.  Shnir,
{\it Magnetic Monopoles} (Springer Berlin Heidelberg New York, 2005).

\bibitem{tHooft:1974kcl}
  G.~'t Hooft,
  ``Magnetic Monopoles in Unified Gauge Theories,''
  Nucl.\ Phys.\ B {\bf 79}, 276 (1974).

\bibitem{Polyakov:1974ek}
  A.~M.~Polyakov,
  ``Particle Spectrum in the Quantum Field Theory,''
  JETP Lett.\  {\bf 20}, 194 (1974)
  [Pisma Zh.\ Eksp.\ Teor.\ Fiz.\  {\bf 20}, 430 (1974)].

\bibitem{Nambu:1961tp}
Y.~Nambu and G.~Jona-Lasinio,
``Dynamical Model of Elementary Particles Based on an Analogy with Superconductivity. 1.,''
Phys.\ Rev.\  {\bf 122} (1961) 345.

\bibitem{Klevansky:1992qe}
S.~P.~Klevansky,
``The Nambu-Jona-Lasinio model of quantum chromodynamics,''
Rev.\ Mod.\ Phys.\  {\bf 64}, 649 (1992).

\bibitem{Finkelstein:1951zz}
R.~Finkelstein, R.~LeLevier, and M.~Ruderman,
``Nonlinear Spinor Fields,''
Phys.\ Rev.\  {\bf 83}, 326 (1951).

\bibitem{Finkelstein:1956}
R. Finkelstein, C. Fronsdal, and P. Kaus,
``Nonlinear Spinor Field,''
Phys.\ Rev.\  {\bf 103}, 1571 (1956).


\bibitem{Deser:1976wq}
S.~Deser,
``Absence of Static Solutions in Source-Free Yang-Mills Theory,''
Phys.\ Lett.\  {\bf 64B}, 463 (1976).

\bibitem{Li:1982gf}
X.~z.~Li, K.~l.~Wang, and J.~z.~Zhang,
``Light Spinor Monopole,''
Nuovo Cim.\ A {\bf 75}, 87 (1983).

\bibitem{Li:1985gf}
K.~L.~Wang and J.~Z.~Zhang,
``The Problem Of Existence For The Fermion - Dyon Selfconsistent Coupling System In A Su(2) Gauge Model,''
Nuovo Cim.\ A {\bf 86}, 32 (1985).

\end{thebibliography}


\section*{Acknowledgements}

This work was supported by Grant No.~BR05236730 in Fundamental Research in Natural Sciences by the Ministry of Education and Science of the Republic of Kazakhstan. V.D. and V.F. also are grateful to the Research Group Linkage Programme of the Alexander von Humboldt Foundation for the support of this research.

\end{document}